\documentclass[11pt,onecolumn]{article}
\usepackage[left=1in,top=1in,right=1in,bottom=1in]{geometry}
\usepackage{graphicx,caption,subfigure,epstopdf,grffile,hyperref,authblk,topcapt,enumerate}
\usepackage{amssymb,amsmath,xcolor}
\usepackage{setspace,lineno} 
\graphicspath{{./Figures/}}	
\newcommand{\Ca}{Ca$^{2+}$}
\newcommand{\Nap}{Na$^+$}
\newcommand{\Kp}{K$^+$}

\DeclareMathOperator{\eps}{\varepsilon}

\title{\vspace{-20pt}Why Pacing Frequency Affects the Production of Early Afterdepolarizations in Cardiomyocytes: An Explanation Revealed by Slow/Fast Analysis of a Minimal Model}
\author{
Theodore Vo\thanks{
		Department of Mathematics, 
		FSU, 
		Tallahassee, FL 32306, USA 
		(\href{mailto:vo@math.fsu.edu}{vo@math.fsu.edu})
		}
\and
Richard Bertram\thanks{
		Department of Mathematics and Programs in Neuroscience and Molecular Biophysics, 
		FSU, 
		Tallahassee, FL 32306, USA 
		(\href{mailto:vo@math.fsu.edu}{bertram@math.fsu.edu})
		}		
}

\begin{document}
\bibliographystyle{vancouver}
\maketitle

\vspace{-10pt}
\begin{abstract}	\label{sec:abstract}
\noindent Early afterdepolarizations (EADs) are pathological voltage oscillations in cardiomyocytes that have been observed in response to a number of pharmacological agents and disease conditions. Phase-2 EADs consist of small voltage fluctuations that occur during the plateau of an action potential, typically under conditions in which the action potential is elongated. Although a single-cell behavior, EADs can lead to tissue-level arrhythmias, including ventricular tachycardia. Much is currently known about the biophysical mechanisms (i.e., the roles of ion channels and intracellular \Ca\ stores) for the various forms of EADs, due partially to the development and analysis of mathematical models. This includes the application of slow/fast analysis, which takes advantage of timescale separation inherent in the system to simplify its analysis. We take this further, using a minimal 3D model to demonstrate that the phase-2 EADs are canards that are formed in the neighborhood of a folded node singularity. This knowledge allows us to determine the number of EADs that can be produced for a given parameter set without performing computer simulations, and provides guidance on parameter changes that can facilitate or inhibit EAD production. With this approach, we demonstrate why periodic stimulation, as would occur in an intact heart, preferentially facilitates EAD production when applied at low frequencies,. We also explain the origin of complex alternan dynamics that can occur with intermediate-frequency stimulation, in which  varying numbers of EADs are produced with each stimulation. These revelations fall out naturally from an understanding of folded node singularities, but are hard or impossible to glean from a knowledge of the biophysical mechanism for EADs alone. Therefore, an understanding of the canard mechanism is a useful complement to an understanding of the biophysical mechanism that has been developed over years of experimental and computational investigations.

\vspace{8pt}\noindent \textbf{Keywords}\qquad canard, action potential, early afterdepolarization, heart, myocyte

\end{abstract}

\section*{Introduction}		\label{sec:intro}
In the normal heart, each heartbeat is associated with an action potential (AP).
The cardiac AP consists of a depolarized phase in which the voltage is elevated; this is associated with transient increased permeability of the cell membrane to Na$^+$ and Ca$^{2+}$. 
The depolarized phase is followed by a repolarization to the resting membrane potential, associated with increased permeability to K$^+$ ions. 
These changes in the membrane potential lead to a sequence of events that result in contraction of the heart muscle, thus allowing for the pumping of blood through the body. 

Early afterdepolarizations (EADs) are pathological voltage oscillations that have been observed in heart muscle cells (cardiomyocytes) during the repolarizing phase of the cardiac AP under conditions in which the AP is elongated. EADs can be induced by hypokalemia \cite{Madhvani2011,Sato2010}, as well as oxidative stress \cite{Xie_LH2008}. They have also been often observed following the administration of drugs that act on \Kp, \Nap, or \Ca\ ion channels such as dofetilide \cite{Guo2007}, {\em dl}-sotalol \cite{Yan2001}, azimilide \cite{Yan2001}, bepridil \cite{Nobe1993,Winslow1986}, isoproterenol \cite{Priori1990,Shimizu1991}, quinidine \cite{Davidenko1989}, and BayK8644 \cite{January1989,Sato2010}. These drug-induced EADs can then lead to ventricular tachyarrhythmias \cite{Asano1997,ElSherif2003, Yan2001}. Genetic defects in \Nap\ and \Kp\ channels that prolong the action potential duration can also lead to an increased rate of EADs and risk of sudden death \cite{Napolitano2005}. 

EADs have been associated with long QT syndrome  \cite{Shimizu1991}, and have long been recognized as a mechanism for the generation of premature ventricular complexes (PVCs) in the electrocardiogram \cite{Shimizu1994}. Different ventricular arrhythmias, including torsade de pointes, are thought to be initiated by PVCs stemming from EADs \cite{Cranefield1991,Lerma2007stochastic,Shimizu1997,Shimizu1991}.
That is, EADs at the myocyte level have been implicated as the primary mechanism promoting arrhythmias at the tissue level in acquired and congenital long-QT syndromes, including polymorphic ventricular tachycardia and ventricular fibrillation \cite{Pogwizd2004,Sanguinetti2006, Yan2001}.

Numerous mathematical models have been constructed at the cellular level to study the genesis of EADs  \cite{Kurata2017,Luo1994b,Sato2010,Tran2009,Zeng1995}. These have confirmed the importance of increased inward \Ca\ current and decreased outward \Kp\ current in the production of EADs. They have also confirmed that reactivation of \Ca\ current is a key element of EAD production \cite{Zeng1995}. The importance of this ``\Ca\ window current" in EAD production was later demonstrated through the use of the Dynamic Clamp technique \cite{Madhvani2011}, which is a hybrid between mathematical modeling and experimentation. Modeling at the tissue level has also been done, in this case to understand EAD propagation, synchronization and the genesis of arrthymia \cite{DeLange2012,Huffaker2004,Sato2009,Vandersickel2014}. These studies demonstrate that EADs at the cellular level can lead to arrhythmias at the tissue level, as has been suggested in experimental studies. 

A useful analysis technique for understanding the behavior of models of excitable systems such as cardiomyocyte models separates system variables into those that change on a fast time scale and those that change on a slow time scale, and then analyzes the two subsystems and their interaction \cite{Bertram2017}. This slow/fast analysis has been used to understand the genesis of EADs, using a 3-variable model in which two variables were treated as ``fast variables" and one variable treated as a ``slow variable". It was shown that EADs can arise via a delayed subcritical Hopf bifurcation of the fast subsystem of variables \cite{Tran2009,Kugler2016}. This explanation, while providing insights, is limited in its descriptive capabilities. For example, it provides limited information on parameter sets for which EADs may occur, and it does not allow one to predict the number of EADs that are produced when they do occur.  

Recently, it was demonstrated that EADs can be attributed to the existence of a folded-node singularity and the accompanying canard orbits \cite{kugler2018}. This was done with the same
three-dimensional model for cardiac action potentials, but now treating one variable as a fast variable and the other two as slow variables. Such a splitting provides the potential for insights that are not available with the 1-slow/2-fast splitting, as is demonstrated in \cite{kugler2018} and in an earlier publication that focused on electrical bursting in pituitary cells \cite{vo2010}. For example, once it is established that the EADs are organized by a folded-node singularity, it is possible to determine regions of parameter space in which EADs can occur \cite{kugler2018}. 

Ventricular cardiomyocytes are, in a physiological setting, subject to periodic stimulation from upstream cardiac cells, originating at the sinoatrial node. Prior experimental and modeling studies have demonstrated that EADs occur more readily at low pacing frequencies than at high frequencies 
\cite{Sato2010,Zeng1995,Damiano1984}. At intermediate forcing frequencies the dynamics are very complex, consisting of alternans with varying numbers of EADs at each stimulus, a behavior described as ``dynamical chaos" \cite{Sato2010,Tran2009}. The primary goal of this article is to provide an understanding for these phenomena. To achieve this, we use the same minimal cardiac action potential model that was developed in \cite{Sato2010} and used recently in \cite{kugler2018}, and apply a 2-slow/1-fast splitting of the model. We demonstrate that the effects of periodic stimulation of the model cell can be understood precisely using the theory of folded-node singularities. In particular, we show that the number of EADs produced by a stimulus depends on where it injects the trajectory into the so-called ``singular funnel", and with this knowledge we demonstrate why low-frequency pacing is expected to yield more EADs than is high-frequency pacing. We also demonstrate the origin of the ``dynamical chaos" that occurs at intermediate-frequency pacing. Finally, we demonstrate why drugs that inhibit the opening of \Kp\ channels facilitate EADs, and why EADs can be induced by hypokalemia \cite{Madhvani2011,Sato2010,Yan2001}.

\section*{Action Potentials and EADs with the Minimal Model}		\label{sec:model}
We study a low-dimensional model for the electrical activity in a cardiomyocyte \cite{Sato2010}, 
\begin{equation} \label{eq:model}
\begin{split}
C_m \frac{dV}{dt} &= -\left( I_{\rm K} + I_{\rm Ca} \right) + I_{\rm sti}, \\
\frac{dn}{dt} &= \frac{n_{\infty}(V)-n}{\tau_n}, \\
\frac{dh}{dt} &= \frac{h_{\infty}(V)-h}{\tau_h},
\end{split}
\end{equation}
where $I_{\rm K}$ is a repolarizing K$^+$ current, $I_{\rm Ca}$ is a depolarizing Ca$^{2+}$ current, and $I_{\rm sti}$ is an external pacmaking stimulus current.
We note that \eqref{eq:model} excludes the depolarizing \Nap\ current since prior studies have found that it has almost no effect on EADs (since it is inactivated during the plateau of the AP).
Here, $V$ is the membrane potential across the cell, $n$ is the activation variable for the K$^+$ channels, and $h$ is the inactivation variable for the L-type Ca$^{2+}$ channels. The ionic currents are described by
\[ I_{\rm K} = g_K n \left( V - V_{\rm K} \right) \quad \text{ and } \quad I_{\rm Ca} = g_{Ca} m_{\infty}(V) h \left( V - V_{\rm Ca} \right), \]
and we set 
\[ I_{\rm sti} =  40 \sum_{k \in \mathbb{N}} \left[ H\left( t - k \cdot {\rm PCL} \right) - H\left( t - (k \cdot {\rm PCL}+1)\right) \right] \]  
where $H(\cdot)$ is the Heaviside function.
That is, the stimulus current provides the system with square wave pulses of $1$~ms duration and $40$~$\mu$A/cm$^2$ amplitude at a frequency set by the pacing cycle length (PCL).  The steady state activation functions are 
\[ x_{\infty}(V) = \frac{1}{1+\exp \left( \frac{V_x-V}{s_x} \right)},  \]
where $x \in \{ m, n \}$, and the steady state inactivation function is 
\[ h_{\infty}(V) = \frac{1}{1+\exp \left( \frac{V-V_h}{s_h} \right)}.  \]
Standard parameter values are listed in Table \ref{tab:params}; these have been tuned so that the model \eqref{eq:model} periodically produces APs with EADs even in the absence of any stimulus current, as in \cite{Sato2010}. 

\begin{table}[ht]
\centering
\topcaption{Parameters definitions and typical values used in the minimal model \eqref{eq:model}.} 
\begin{tabular}{|c|c|l|}
\hline
Parameter & Value & Definition \\
\hline 
$C_m$ & 0.5 $\mu$F/cm$^2$ & Membrane capacitance \\
$g_{Ca}$ & 0.025~mS/cm$^2$ & Maximal conductance of L-type Ca$^{2+}$ channels \\
$g_K$ & 0.04~mS/cm$^2$ & Maximal conductance of K$^{+}$ channels \\
$V_{Ca}$ & 100~mV & Reversal potential for Ca$^{2+}$ \\
$V_{K}$ & -80~mV & Reversal potential for K$^{+}$ \\
$\tau_n$ & 300~ms & Time constant for activation of K$^{+}$ channels \\
$\tau_h$ & 80~ms & Time constant for activation of Ca$^{+}$ channels \\
$V_m$ & -35~mV & Voltage value at midpoint of $m_{\infty}(V)$ \\
$s_m$ & 6.24~mV & Slope parameter of $m_{\infty}(V)$ \\
$V_n$ & -40~mV & Voltage value at midpoint of $n_{\infty}(V)$ \\
$s_n$ & 5~mV & Slope parameter of $n_{\infty}(V)$ \\
$V_h$ & -20~mV & Voltage value at midpoint of $s_{\infty}(V)$ \\
$s_h$ & 8.6~mV & Slope parameter of $s_{\infty}(V)$ \\
\hline
\end{tabular}
\label{tab:params}
\end{table}

The model cell \eqref{eq:model} exhibits two distinct AP morphologies: regular APs and APs with EADs. 
For the remainder of the article, we use the Farey sequence notation, $1^s$, to denote a single large-amplitude AP with $s$ small-amplitude EADs during the repolarizing phase. 
Thus, a regular AP is denoted $1^0$ and an AP with 2 EADs is denoted $1^2$.
More complicated rhythms are described using concatenations of these Farey sequences. 
For instance, a rhythm that periodically exhibits three regular APs followed by a single AP with 2 EADs is denoted $(1^0)^3 (1^2)$.

\subsection*{Action Potential Duration and Number of EADs Increases with PCL}		\label{subsec:bifurcation}
\begin{figure}[ht]
\centering
\includegraphics[width=5in]{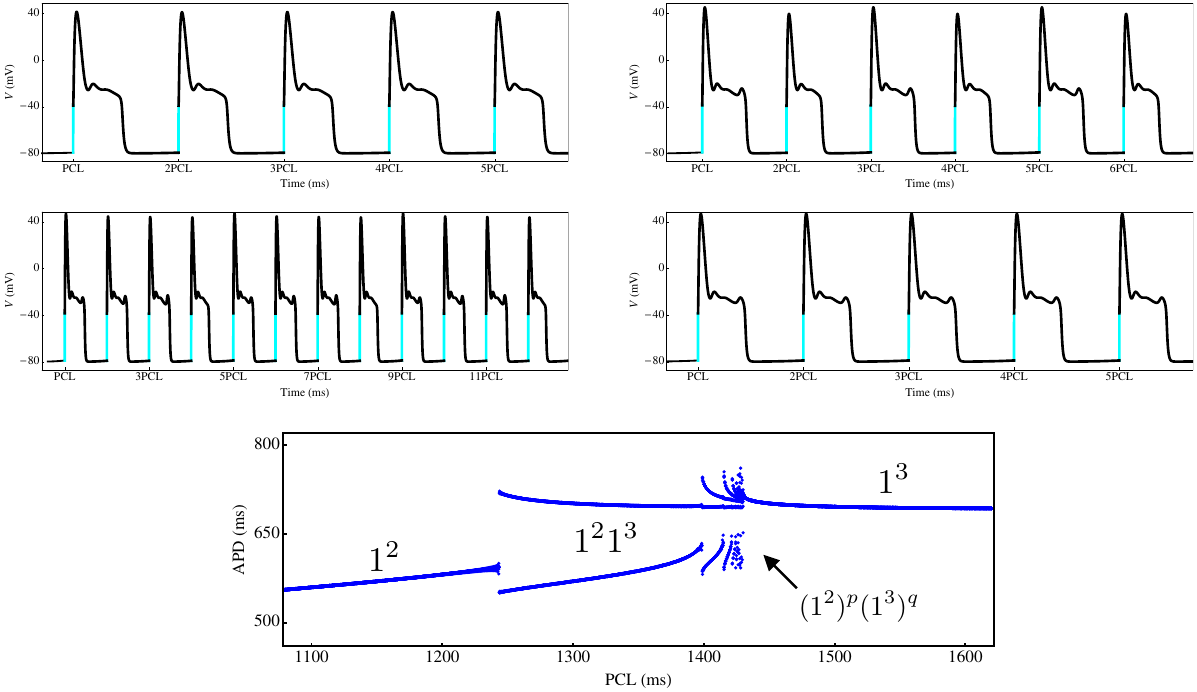}
\put(-368,201){(a)}
\put(-182,201){(b)}
\put(-368,138){(c)}
\put(-182,138){(d)}
\put(-300,72){(e)}
\caption{Dynamics of the model cardiomyocyte \eqref{eq:model} under variations in the PCL. In (a)--(d), the stimulus pulse is `on' during the cyan segments. The attractor of the cell shows (a) $1^2$ APs with EADs for ${\rm PCL} = 1200~$ms, (b) $1^2 1^3$ alternans for ${\rm PCL}=1300~$ms, (c) $1^2 (1^3)^3$ APs with EADs for ${\rm PCL} = 1420$~ms, and (d) $1^3$ APs with EADs for ${\rm{PCL} = 1500}$~ms. (e) APD versus PCL bifurcation diagram. There is an intermediate band of PCLs ($1240$~ms $\lesssim {\rm PCL} \lesssim$ $1435~$ms) over which the attractor has complex EAD signature.}
\label{fig:restitution}
\end{figure}
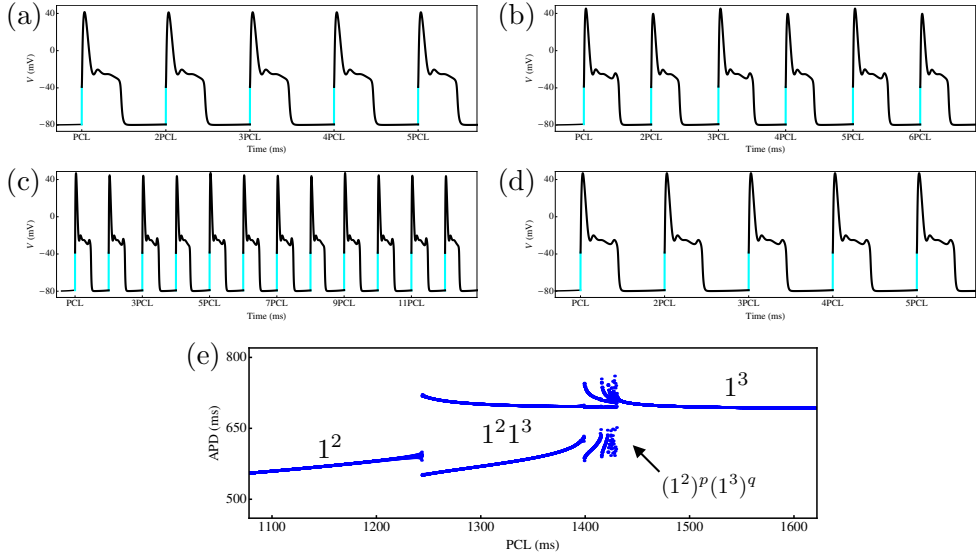

The model cell \eqref{eq:model} is entrained to the periodic stimulus; for the parameter set in Table \ref{tab:params}, the cell exhibits $1^s$ impulses with period set by the PCL. For small PCLs (i.e., high-frequency pulsing), the attractor is a $1^2$ rhythm (Fig. \ref{fig:restitution}(a)). For intermediate PCLs ($1240$~ms $\lesssim {\rm PCL} \lesssim$ $1435~$ms), the cell exhibits complex EAD activity, including $1^2 1^3$ alternans (Fig. \ref{fig:restitution}(b)) and $1^2 (1^3)^3$ rhythms (Fig. \ref{fig:restitution}(c)). For large PCLs (i.e., low-frequency pulsing), the cell is in a $1^3$ state (Fig. \ref{fig:restitution}(d)). 

We summarize the behaviour of the model cell and its response to periodic stimulation at various frequencies, by constructing a bifurcation diagram (Fig. \ref{fig:restitution}(e)). To do this, we used a dynamic restitution protocol \cite{Koller1998} in which the cell was paced at a fixed PCL until steady-state was reached, after which the action potential duration (APD) and PCL were recorded. We took the APD to be the amount of time the cell spends with $V > -70~$mV. With this choice of restitution protocol, the PCL is the sum of the APD and the diastolic interval, so our bifurcation diagram encodes the restitution curves (i.e., the plot of the APD as a function of the diastolic interval has the same qualitative features as shown in Fig. \ref{fig:restitution}(e)).  

The bifurcation diagram shows that the periodic stimulation elicits three types of behaviour. 
For high- and low-frequency stimulation, the model cell is in a purely $1^2$ or $1^3$ state, respectively. 
In the intermediate-frequency forcing range, the model cell has complex signature of the form $(1^2)^p (1^3)^q$, where $p$ and $q$ are integers. We observe that the AP signature becomes more complicated near the transition to the $1^3$ state. 
This increasing complexity of the AP signature near a transition is robust; it occurs for a wide range of $g_K$ and $g_{Ca}$ in \eqref{eq:model} and has also been observed in other forced conductance-based cardiomyocyte models \cite{Sato2010,Tran2009}.  

Now that we have demonstrated the rich variety of dynamics present in the minimal model \eqref{eq:model},
we next investigate the dynamical mechanisms that underlie the observed rhythms.  
We use Geometric Singular Perturbation Theory \cite{Fenichel1979,Jones1995} as the basis of our analysis. 

\section*{EADs Arise From Canard Dynamics}		\label{sec:local}
In this section, we show that the dynamical mechanisms responsible for the EADs are canards. To facilitate the analysis, we consider \eqref{eq:model} with no stimulus input.  A similar demonstration was provided by \cite{kugler2018}, but we elaborate on how the EADs emerge as the cell capacitance is increased from 0, (i.e., moving the system away from the singular limit), and we demonstrate how the underlying rotational sectors determine the number and duration of EADs. We first show that the model has a slow/fast structure. We use this slow/fast splitting to identify the geometric cast of characters involved in producing APs and EADs.  We then demonstrate that folded node canards generate EADs, and these canards are robust in parameters. Finally, we demonstrate how drugs that inhibit \Kp\ channels or a hypokalemic environment can facilitate EAD production.

\subsection*{The dynamics evolve over multiple timescales} 	\label{subsec:timescales}
A key observation is that the dynamics of the cell evolve over multiple timescales, with slow depolarized/hyperpolarized epochs interspersed with rapid transitions between them. We formally show this multi-timescale structure by introducing dimensionless variables, $v$ and $t_s$, via the rescalings 
\[ v = \frac{V}{k_V} \quad \text{ and } \quad t_s = \frac{t}{k_t}, \]
where $k_V$ and $k_t$ are reference voltage and timescales, respectively. 
With these rescalings, the minimal model \eqref{eq:model} becomes
\begin{equation} \label{eq:dimless}
\begin{split}
\eps \frac{dv}{dt_s} &=  - \left( \overline{g}_K n \left( v-\overline{V}_{\rm K} \right) + \overline{g}_{Ca} m_{\infty}(v) h \left( v-\overline{V}_{\rm Ca} \right) \right) \equiv f(v,n,h), \\
\frac{dn}{dt_s} &= \frac{k_t}{\tau_n} \left( n_{\infty}(v) - n \right) \equiv g_1(v,n), \\
\frac{dh}{dt_s} &= \frac{k_t}{\tau_h} \left( h_{\infty}(v) - h \right) \equiv g_2(v,h), 
\end{split}
\end{equation}
where $\overline{g}_u = \frac{g_u}{g_{\rm ref}}$ for $u \in \{ \rm K,Ca \}$ denotes the dimensionless conductances with reference conductance $g_{\rm ref}$, $\overline{V}_u = \frac{V_u}{k_V}$ for $u \in \{ \rm K,Ca \}$ denotes the dimensionless reversal potentials, and $0 < \eps = \frac{C_m / g_{\rm ref}}{k_t} \ll 1$ is the ratio of the voltage timescale ($C_m/g_{\rm ref}$) to the reference timescale.
The benefit of recasting the model in the dimensionless form \eqref{eq:dimless} is that it reveals the typical timescales in the model. 
The voltage variable is fast with a timescale of $\frac{C_m}{g_{\rm ref}} \approx 5$~ms for $C_m = 0.5~\mu$F/cm$^2$ and $g_{\rm ref}=0.1$~mS/cm$^2$. 
The activation variable, $n$, for the \Kp\ channels is slow with timescale $\tau_n = 80$~ms, and the inactivation variable, $h$, for the L-type \Ca\ channels is superslow with timescale $\tau_s = 300~$ms. 
Thus, the system \eqref{eq:dimless} is a three-timescale problem. 

One effective approach to the analysis of multiple-timescale problems, as pioneered in the neuroscience context in \cite{Rinzel1987}, is Geometric Singular Perturbation Theory (GSPT).
The idea of GSPT is to decompose a slow/fast system into lower dimensional slow and fast subsystems, analyze these simpler subsystems, and combine their information in order to understand the origin and properties of the dynamics of the original model. 
However, the GSPT approach is currently limited to two-timescale (i.e., slow/fast) problems. 
In three-timescale systems such as \eqref{eq:dimless}, a choice is usually made: to either group $v$ and $n$ together as `fast', or to group $n$ and $h$ together as `slow'. 

Prior studies of the minimal model chose to group $v$ and $n$ together as fast, whilst using $h$ as the sole slow variable \cite{Sato2010}. 
In this 1-slow/2-fast approach, the EADs arise because the depolarized steady state of the $(v,n)$ subsystem loses stability via a Hopf bifurcation (with respect to $h$) leading to oscillations which are destroyed at a homoclinic bifurcation \cite{Sato2009,Tran2009,Xie2007}. 
Whilst this mechanism is consistent with the {\em in-vitro} and {\em in-silico} observations that the EADs appear irregularly under periodic stimulation, it does not provide insight into how many EADs should be observed or why the number of EADs change with the PCL. 
Here, we take the alternative approach and treat $v$ as the only fast variable, whilst grouping $n$ and $h$ together as slow. 
We will show that this 2-slow/1-fast approach allows us to predict the maximal number of EADs that can be generated, and explain why the number of EADs changes with the PCL.

\subsection*{Underlying geometric structure} 	\label{subsec:gspt}
We now identify the geometric features that organize the EADs and APs. 
We begin by reformulating \eqref{eq:dimless} in terms of the fast time, $t_f = \frac{1}{\eps} t_s$, which gives
\begin{equation} \label{eq:fast}
\begin{split}
\frac{dv}{dt_f} &= f(v,n,h), \\
\frac{dn}{dt_f} &= \eps g_1(v,n), \\
\frac{dh}{dt_f} &= \eps g_2(v,h).
\end{split}
\end{equation}
System \eqref{eq:fast} is equivalent to \eqref{eq:dimless} in the sense that solutions of both systems trace out the same paths in the $(v,n,h)$ phase space, just at different speeds. 
We have seen that the dynamics of \eqref{eq:model} alternate between slow and fast epochs. 
The rapid transitions between depolarized and repolarized phases are approximated by solutions of the 1D {\em fast subsystem}
\begin{equation} \label{eq:layer}
\begin{split}
\frac{dv}{dt_f} &= f(v,n,h), \\
\frac{dn}{dt_f} &= 0, \\
\frac{dh}{dt_f} &= 0,
\end{split}
\end{equation}
which is the approximation of \eqref{eq:dimless} in which the slow variables move so slowly that they are fixed. 
(The fast subsystem is obtained by taking the singular limit $\eps \to 0$ in \eqref{eq:fast}.) 
Similarly, the slow depolarized/repolarized segments of the dynamics are approximated by solutions of the 2D {\em slow subsystem}
\begin{equation} \label{eq:reduced}
\begin{split}
0 &= f(v,n,h), \\
\frac{dn}{dt_s} &= g_1(v,n), \\
\frac{dh}{dt_s} &= g_2(v,h),
\end{split}
\end{equation}
which is the approximation of \eqref{eq:dimless} in which the fast voltage variable moves so rapidly that it (i) has already reached steady state and (ii) instantly adjusts to any changes in the slow gating dynamics. 
(The slow subsystem is obtained by taking the singular limit $\eps \to 0$ in \eqref{eq:dimless}.) 
Recall that the idea of GSPT is to analyze the 1D fast and 2D slow subsystems, and combine their information in order to understand the origin and properties of the dynamics in the full 3D system.

\begin{figure}[ht]
\centering
\includegraphics[width=5in]{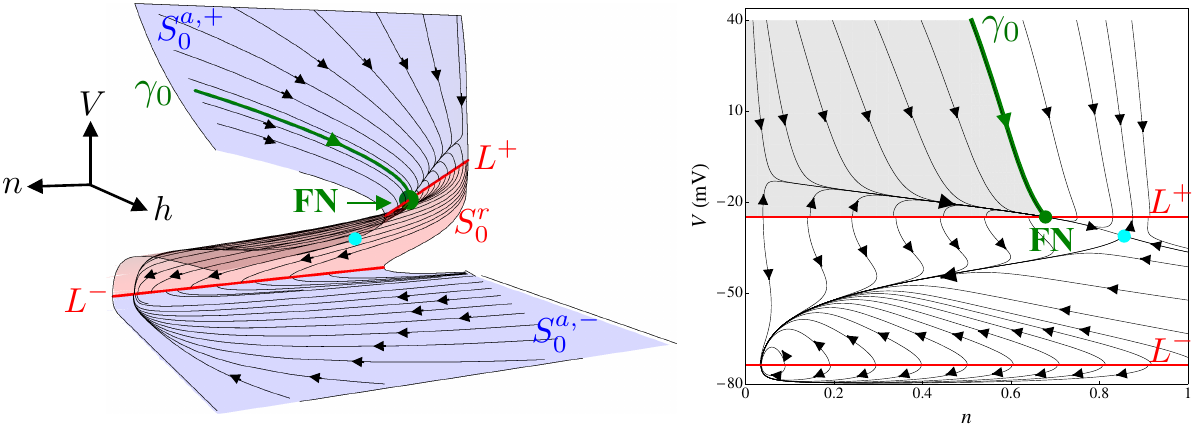}
\put(-360,120){(a)}
\put(-162,120){(b)}
\caption{Geometric structure of the cardiomyocyte model \eqref{eq:model} for the parameter set in Table \ref{tab:params}. (a) The outer attracting sheets (blue surfaces) of the critical manifold are separated from the middle repelling sheet (red surface) by the (red) fold curves, $L^{\pm}$. The slow flow (given by \eqref{eq:slowprojection}; black curves) is directed towards the folds. There is a folded node (green marker) on $L^+$ with singular strong canard, $\gamma_0$ (green trajectory). The full system equilibrium (cyan marker) is a saddle. (b) Projection into the $(V,n)$ plane. The funnel region (gray) for trajectories that enter FN is enclosed by $L^+$ and $\gamma_0$.}
\label{fig:slowflow}
\end{figure}

We begin with linear stability analysis of the 1D fast subsystem \eqref{eq:layer}.
The equilibria, $S_0$, of \eqref{eq:layer} form a cubic-shaped surface (in the $(v,n,h)$ space) called the critical manifold
\begin{equation} \label{eq:criticalmanifold} 
S_0 = \left\{ (v,n,h) : f(v,n,h) = 0 \right\} = \left\{ (v,n,h) : h = h_S(v,n) = - \frac{\overline{g}_{K} n(v-\overline{V}_K)}{\overline{g}_{Ca} m_{\infty}(v) (v-\overline{V}_{Ca}) } \right\}.
\end{equation}
The outer sheets are stable and the middle sheet is unstable; these are separated by curves, $L^{\pm}$, of points corresponding to fold bifurcations of \eqref{eq:layer}
\begin{equation} \label{eq:foldcurves} 
L^{\pm} = \left\{ (v,n,h) \in S_0 : \frac{\partial f}{\partial v} = 0 \right\}.
\end{equation}
For the cardiomyocyte model, the fold conditions \eqref{eq:foldcurves} reduce to a set of lines on $S_0$ at constant voltage values (Fig.~\ref{fig:slowflow}; red curves); $L^+$ denotes the fold curve at a depolarized voltage level, and $L^-$ denotes the fold curve at a hyperpolarized voltage that is the firing threshold. 
We note that the $V$-axis is also a fold curve (see `\nameref{subsec:twoparam}' section).

From the linear stability analysis, we conclude that most solutions of \eqref{eq:layer} end up on either the depolarized attracting sheet, $S_0^{a,+}$, or the hyperpolarized attracting sheet, $S_0^{a,-}$.
Once trajectories reach one of these sheets, the slow dynamics dominate the evolution and the appropriate approximating system is the slow subsystem \eqref{eq:reduced}. 
The algebraic equation in \eqref{eq:reduced} constrains the phase space to the critical manifold, whilst the differential equations describe the slow motions along $S_0$.
Thus, the geometric singular perturbation analysis partitions the phase space into the fast dynamics away from the critical manifold together with the slow dynamics on the critical manifold. 
The critical manifold itself is the interface where the fast and slow subsystems interact.

For the slow evolution on $S_0$, we have differential equations to describe the motions of $n$ and $h$, whilst the algebraic equation implicitly describes the associated motions in $v$ (slaved to $S_0$; Fig.~\ref{fig:slowflow} black curves). 
To obtain an explicit description of the $v$-motions, we differentiate $f(v,n,h)=0$ with respect to the slow time, $t_s$, and use the graph representation of the critical manifold given in \eqref{eq:criticalmanifold}. 
This gives 
\begin{equation}	\label{eq:slowprojection}
\begin{split} 
\frac{dv}{dt_s} &=-\left( \frac{\partial f}{\partial v}\right)^{-1} \left(\frac{\partial f}{\partial n} g_1 + \frac{\partial f}{\partial s} g_2 \right), \\
\frac{dn}{dt_s} &= g_1,
\end{split}
\end{equation}
where $h$ has been replaced by $h_S(v,n)$.
We stress that \eqref{eq:slowprojection} is equivalent to \eqref{eq:reduced}; we have simply incorporated the restriction to $S_0$ explicitly by setting $h = h_S(v,n)$. 
In this formulation, it becomes clear that the slow flow is singular along the fold curves, $L^\pm$, where $\frac{\partial f}{\partial v} = 0$.
To deal with this finite-time blow-up of solutions, we perform the time rescaling $dt_s = -\frac{\partial f}{\partial v} dt_d$, which transforms the slow system \eqref{eq:slowprojection} to the {\em desingularized system},
\begin{equation}	\label{eq:desingularized}
\begin{split} 
\frac{dv}{dt_d} &=\frac{\partial f}{\partial n} g_1 + \frac{\partial f}{\partial s} g_2, \\
\frac{dn}{dt_d} &= -\left( \frac{\partial f}{\partial v}\right) g_1,
\end{split}
\end{equation}
where again, $h = h_S(v,n)$. 
In this setting, the finite-time singularities of \eqref{eq:slowprojection} along the fold curves have been transformed into nullclines of \eqref{eq:desingularized}.
Since the transformation that led to the desingularized system is phase space-dependent, some care must be taken when comparing trajectories of the desingularized system \eqref{eq:desingularized} with those of the true slow subsystem \eqref{eq:slowprojection}.
On the attracting sheets, $S_0^{a,\pm}$, the flow of \eqref{eq:desingularized} is topologically equivalent to the flow of \eqref{eq:slowprojection} since $\frac{\partial f}{\partial v}<0$ (and hence $t_s$ and $t_d$ have the same sign). 
On the repelling sheet, $S_0^r$, the flow of \eqref{eq:desingularized} is in the opposite direction to the flow of \eqref{eq:slowprojection} since $\frac{\partial f}{\partial v}>0$ (and hence $t_s$ and $t_d$ have opposite signs).

With this relation between the slow and desingularized systems in mind, we now analyze the desingularized system in order to learn about the dynamics of the slow subsystem. 
The desingularized system possesses two types of equilibria or singularities. 
Ordinary singularities are isolated points such that $\{ g_1 = g_2 = 0 \}$, and correspond to true equilibria of the desingularized system \eqref{eq:desingularized}, of the slow subsystem \eqref{eq:slowprojection}, and of the original model \eqref{eq:model}.
For the parameter set in Table \ref{tab:params}, there is an ordinary singularity on $S_0^r$ (Fig. \ref{fig:slowflow}; cyan marker), corresponding to a saddle equilibrium.

Folded singularities, $M$, are isolated points on $L^{\pm}$ where the right-hand-side of the $v$-equation in \eqref{eq:desingularized} is zero, i.e.,
\begin{equation} \label{eq:foldedsing}
M = \left\{ (v,n,h) \in L^{\pm} : \frac{\partial f}{\partial n} g_1 + \frac{\partial f}{\partial h} g_2  =0 \right\}.
\end{equation}
Folded singularities correspond to equilibria of \eqref{eq:desingularized}, however, they are not equilibria of the slow subsystem \eqref{eq:slowprojection} or the original model \eqref{eq:model}. 
Instead, they are points where both the numerator and denominator of the right-hand-side of the $v$-equation in \eqref{eq:slowprojection} vanish at the same time, so there may be a cancellation of a simple zero. 
This allows the possibility of solutions of the slow flow to cross the fold curves (via the folded singularity) with finite speed and move from an attracting sheet to the repelling sheet (or vice versa). 
Such solutions are called {\em singular canards} \cite{Szmolyan2001}, and play important roles in applications. We refer to \cite{Desroches2012,Kuehn2015} for extensive overviews of applications of folded singularities and canards in chemical, neural, and engineering contexts. 

Folded singularities are classified as equilibria of the desingularized system. 
A folded singularity with real eigenvalues of the same sign is a folded node; with real eigenvalues of opposite signs it is a folded saddle; and with complex conjugate eigenvalues it is a folded focus. 
Folded nodes and folded saddles possess singular canards, whereas folded foci do not. 
The cardiomyocyte model possesses a folded node on $L^+$ for the standard parameter set (Fig. \ref{fig:slowflow}; green marker).

\subsection*{EADs originate from a folded node} 	\label{subsec:mmos}

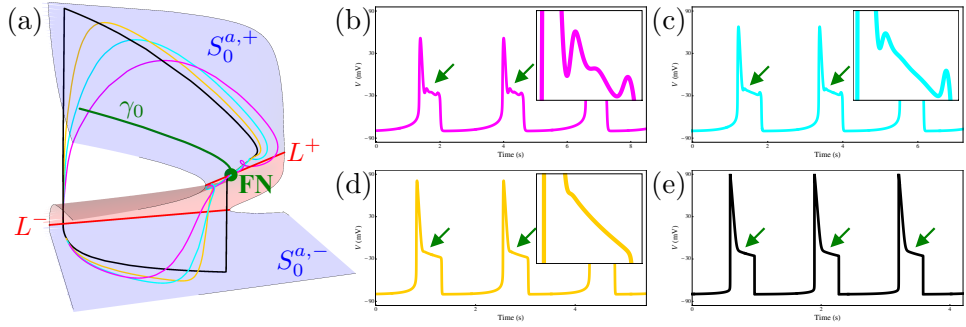
\begin{figure}[ht]
\centering
\includegraphics[width=5in]{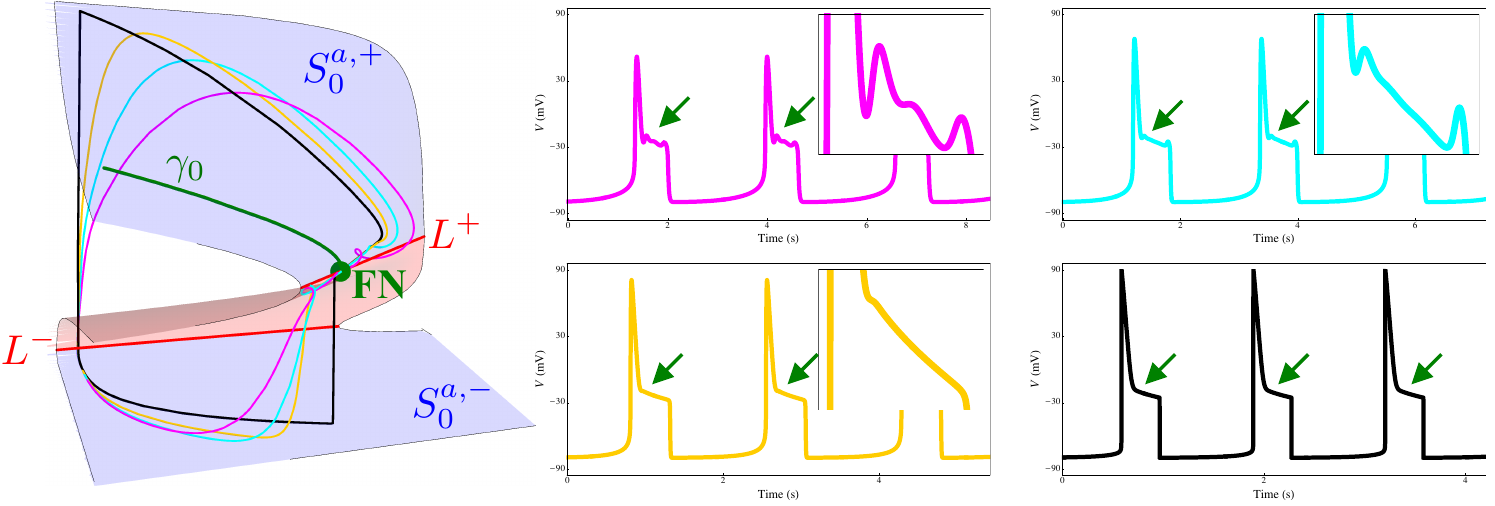}
\put(-364,112){(a)}
\put(-240,112){(b)}
\put(-120,112){(c)}
\put(-240,50){(d)}
\put(-120,50){(e)}
\caption{Origin of the EADs near the folded node (green marker) for the standard parameter set. (a) Singular (black) and nonsingular (magenta, cyan, and yellow) $1^3$ attractor compared to the critical manifold. All orbits enter the depolarized sheet, $S_0^{a,+}$, inside the funnel enclosed by the singular strong canard $\gamma_0$ (green curve) and $L^+$ (red curve). The corresponding voltage time series are shown for (b) $C_m = 0.5~\mu$F/cm$^2$ (magenta), (c) $C_m = 0.25~\mu$F/cm$^2$ (cyan), (d) $C_m = 0.1~\mu$F/cm$^2$ (yellow), and (e) $C_m = 0~\mu$F/cm$^2$ (black).}
\label{fig:epsunfolding}
\end{figure}

We now demonstrate the origin of EADs in terms of the geometric structures identified in the prior section `\nameref{subsec:gspt}'.
To motivate this, we first take a $1^3$ attractor of \eqref{eq:model} (without periodic stimulation) and compare it to the critical manifold in the $(V,n,h)$ phase space (Fig. \ref{fig:epsunfolding}(a); magenta curve). The three EADs can be seen as small loops in the magenta trajectory about the upper fold curve, $L^+$. 
We observe that (i) the EADs are localized to the neighbourhood of the folded node;
(ii) by decreasing $\eps$, or $C_m$, the EADs decrease in amplitude (compare curves of different colors in Fig. \ref{fig:epsunfolding}); 
(iii) by decreasing $\eps$, the location in phase space where the trajectory transitions from a depolarized state to a hyperpolarized state converges to the folded node. 
These observations lead us to conjecture that the EADs observed for $0< \eps \ll 1$ arise from the folded node itself. 

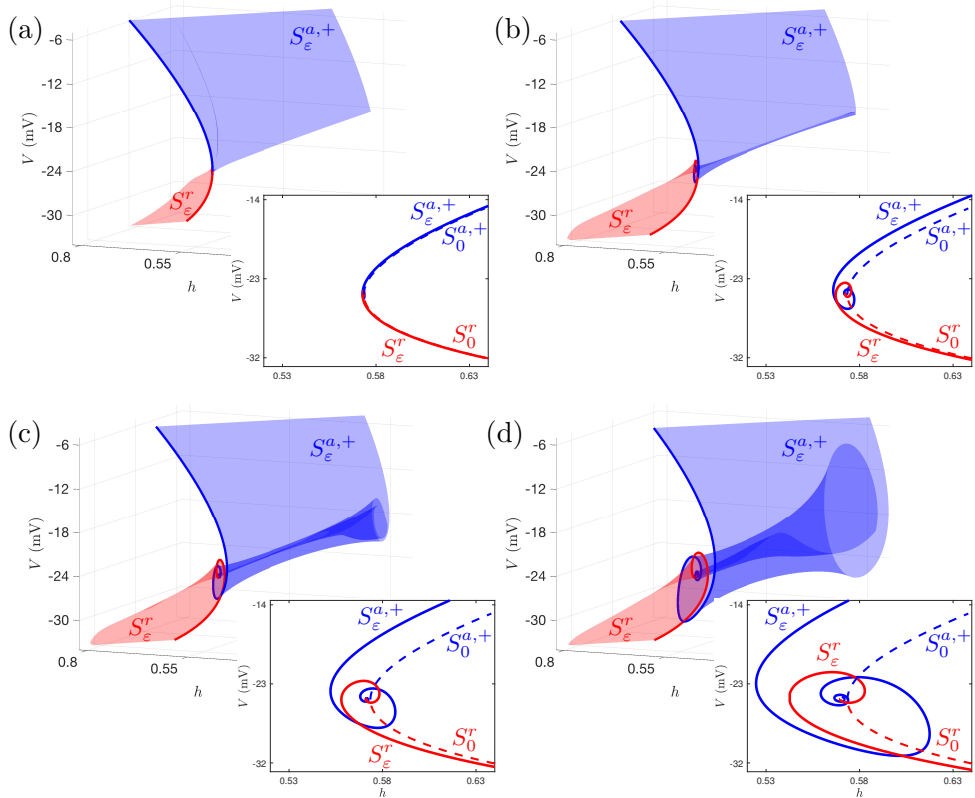
\begin{figure}[ht]
\centering
\includegraphics[width=5in]{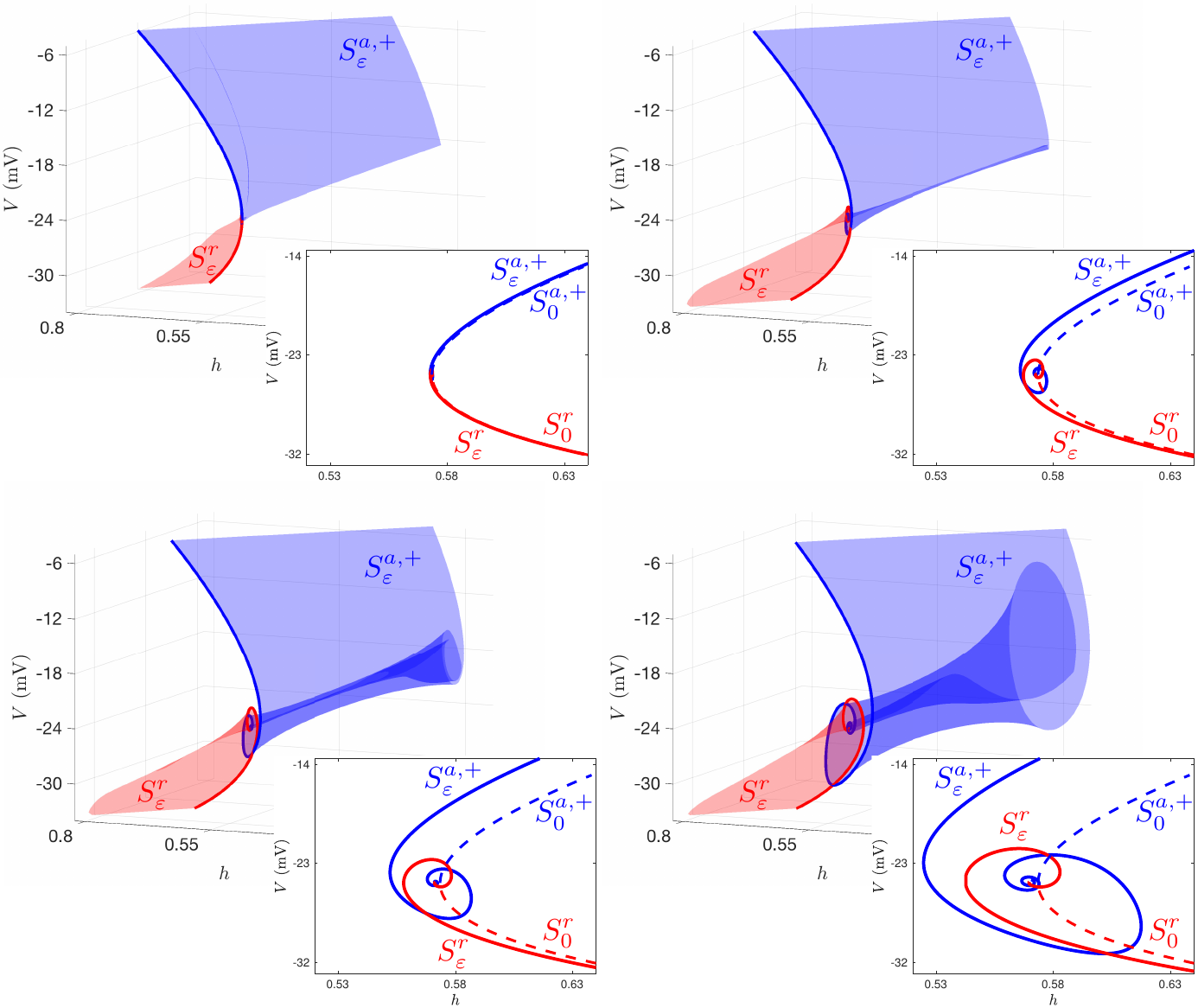}
\put(-366,288){(a)}
\put(-182,288){(b)}
\put(-366,135){(c)}
\put(-185,135){(d)}
\caption{Attracting (blue) and repelling (red) slow manifolds, $S_{\eps}^{a,+}$ and $S_{\eps}^r$,  for (a) $C_m=0.01~\mu$F/cm$^2$, (b) $C_m = 0.1~\mu$F/cm$^2$, (c) $C_m = 0.25~\mu$F/cm$^2$, and (d) $C_m = 0.5~\mu$F/cm$^2$. The twisting of the slow manifolds produces the EADs. Insets: intersections of $S_{\eps}^{a,+}$ (solid blue) and $S_{\eps}^{r}$ (solid red) with $\Sigma$. Also shown for comparison are the intersections of $S_0^{a,+}$ (dashed blue) and $S_0^r$ (dashed red) with $\Sigma$. The folded node is at the intersection of the dashed blue and dashed red curves.}
\label{fig:slowmans}
\end{figure}

How do the small oscillations emerge from the folded node? 
To answer this, we examine how the sheets, $S_0^{a,+}$ and $S_0^r$, of the critical manifold persist for small and nonzero $\eps$. 
As $\eps$ is increased away from zero, the attracting and repelling sheets, $S_0^{a,+}$ and $S_0^r$, perturb to attracting and repelling slow manifolds, $S_{\eps}^{a,+}$ and $S_{\eps}^r$, respectively \cite{Fenichel1979,Jones1995}. 
These slow manifolds are the surfaces to which the slow segments of trajectories of \eqref{eq:model} are slaved.
Both $S_{\eps}^{a,+}$ and $S_{\eps}^r$ are small and regular perturbations of $S_0^{a,+}$ and $S_0^r$, except in the neighbourhood of the folded node, where they instead twist around a common axis of rotation \cite{Szmolyan2001,Wexy2005}. The axis of rotation corresponds to the weak eigendirection of the folded node.
The twisted slow manifolds are shown in Fig. \ref{fig:slowmans} for various perturbations, corresponding to the $C_m$ values used in Fig. \ref{fig:epsunfolding}.
(For the purposes of visualization, the slow manifolds have only been computed up to a plane, $\Sigma$, passing through the folded node. The method of computation is detailed in \cite{Desroches2008}.)
The twisting of the slow manifolds (and the slow flow on them) is confined to an $\mathcal{O}\left(\sqrt{\eps} \right)$ neighbourhood of the folded node \cite{Brons2006}. 
Thus, the EADs arise from locally twisted slow manifolds around the folded node. This can be seen in the 
insets of Fig. \ref{fig:slowmans}, where the folded node is the intersection of the dashed blue curve (intersection of
 $S_0^{a,+}$ with $\Sigma$) and the dashed red curve (intersection of $S_0^{a,-}$ with $\Sigma$).

\subsection*{Canards organize the EADs} 	\label{subsec:sectors}

\begin{figure}[ht]
\centering
\includegraphics[width=5in]{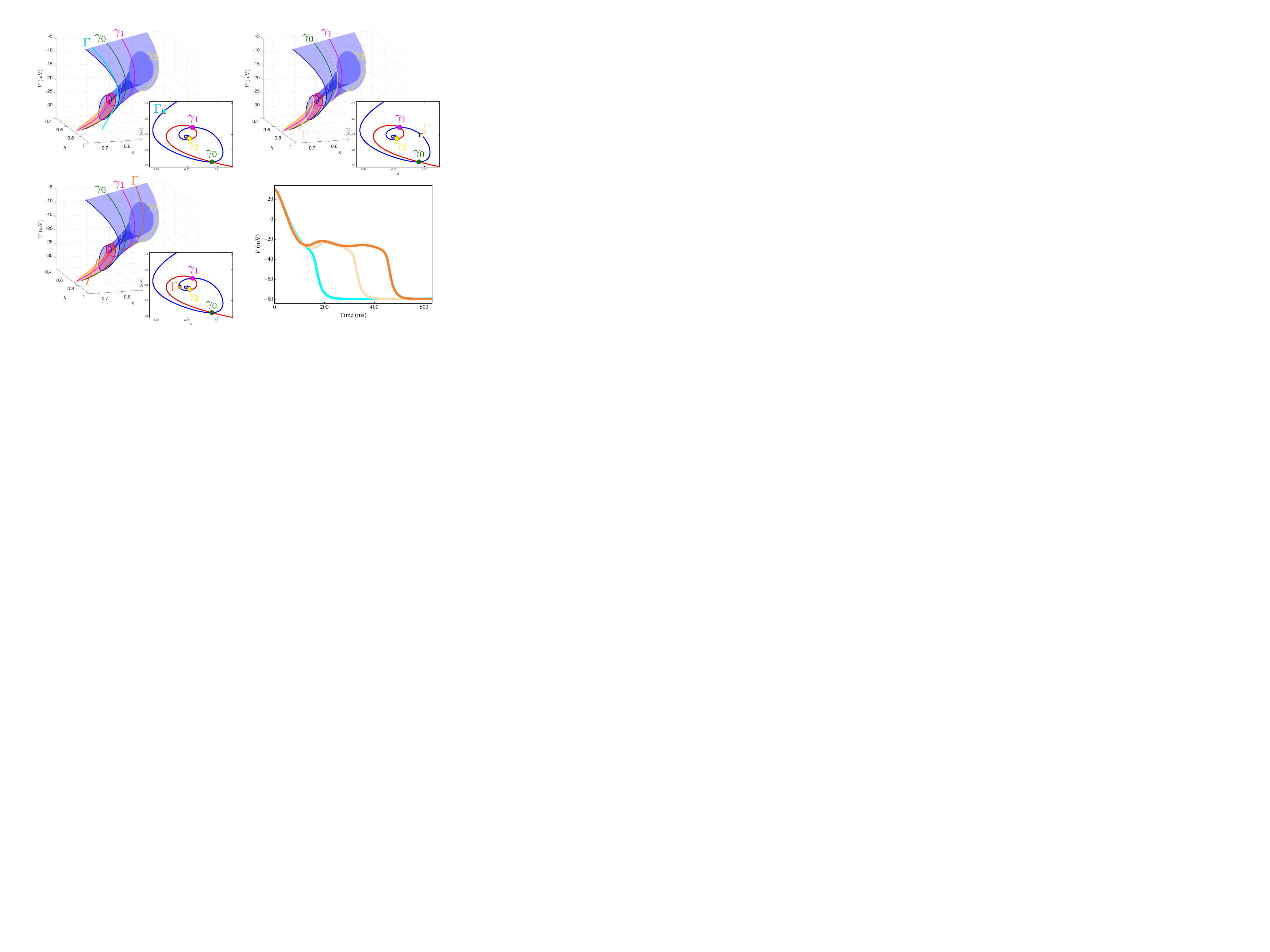}
\put(-366,256){(a)}
\put(-182,258){(b)}
\put(-366,122){(c)}
\put(-182,122){(d)}
\caption{Organization of the EADs by maximal canards for the standard parameter set. Only the first three maximal canards, $\gamma_0$ (green), $\gamma_1$ (magenta) and $\gamma_2$ (yellow), are shown. (a) A solution ($\Gamma$; cyan) outside the rotational sectors has no EADs. (b) A solution ($\Gamma$; beige) in the sector between $\gamma_0$ and $\gamma_1$ exhibits 1 EAD. (c) A solution ($\Gamma$; orange) in the sector between $\gamma_1$ and $\gamma_2$ exhibits 2 EADs. (d) Corresponding time series, showing a regular AP (cyan), an AP with 1 EAD (beige), and an AP with 2 EADs (orange).}
\label{fig:sectors}
\end{figure}

The local twisting of the slow manifolds results in a finite number of intersections between $S_{\eps}^{a,+}$ and $S_{\eps}^r$, called {\em maximal canards}.
For the standard parameter set, we find that there are 5 maximal canards. 
The outermost, $\gamma_0$, is called the {\em maximal strong canard} and is the phase space boundary between those trajectories that exhibit EADs near the folded node and those that do not (Fig. \ref{fig:sectors}). 
That is, any solution of the cardiomyocyte model \eqref{eq:model} with initial condition to the left of $\gamma_0$ in Fig. \ref{fig:sectors} is a regular $1^0$ AP (Fig. \ref{fig:sectors}(a) and (d); cyan curves).
Any solution with initial condition between $\gamma_0$ and the secondary maximal canard $\gamma_1$ executes 1 EAD in the neighbourhood of the folded node (Fig. \ref{fig:sectors}(b) and (d); beige curves). 
Any solution with initial condition enclosed by the secondary canards $\gamma_1$ and $\gamma_2$ exhibits 2 EADs around the folded node (Fig. \ref{fig:sectors}(c) and (d); brown curves). 
In general, an orbit in the sector between the maximal secondary canards $\gamma_{k-1}$ and $\gamma_k$ will execute $k$ EADs.
The innermost maximal canard, $\gamma_w$, is called the {\em maximal weak canard} and is the axis of rotation for both the slow manifolds and the other maximal canards.  
Thus, the maximal canards organize the EADs in phase space; the path taken by the trajectory relative to the maximal canards determines the number of EADs produced.

\subsection*{Folded Node and EAD Dynamics Are Robust} 	\label{subsec:twoparam}

Given that the EADs arise from canard dynamics due to twisted slow manifolds around a folded node, is it possible to predict the number of maximal canards and associated EADs?
The answer is `yes', and it is encoded in the strong and weak eigenvalues, $\lambda_s < \lambda_w <0$, of the folded node (when considered as an equilibrium of the desingularized system). 
Let $\mu = \frac{\lambda_w}{\lambda_s}$ denote the eigenvalue ratio. 
Then, provided $\eps$ is sufficiently small and $\mu \gg \sqrt{\eps}$, the maximal number, $s_{\max}$, of EADs around the folded node  is 
\begin{equation} \label{eq:smax} s_{\max} = \lfloor \frac{\mu+1}{2\mu} \rfloor, \end{equation}
where $\lfloor \frac{\mu+1}{2\mu} \rfloor$ denotes the greatest integer less than or equal to $\frac{\mu+1}{2\mu}$ \cite{Brons2006,Szmolyan2001}. 
The corresponding number of maximal canards is $s_{\max}+1$. 
For the folded node discussed in Figs. \ref{fig:slowflow} -- \ref{fig:sectors}, we find $\mu \approx 0.13$ so that the maximal number of EADs that can be observed is $s_{\max} = 4$, and there are 5 maximal canards (consistent with Fig. \ref{fig:slowmans}). 

Not only does the formula \eqref{eq:smax} predict the number of EADs, it also predicts how the number of EADs changes with parameters. 
Bifurcations of maximal canards occur whenever $\mu^{-1}$ passes through an odd integer value \cite{Wexy2005}. 
That is, if the system parameters are varied so that $\mu^{-1}$ increases through $3$, then $s_{\max}$ increases from $1$ to $2$.
If the system parameters are varied so that $\mu^{-1}$ increases through $5$, then $s_{\max}$ increases from $2$ to $3$, and so on. 

\begin{figure}[ht]
\centering
\includegraphics[width=3in]{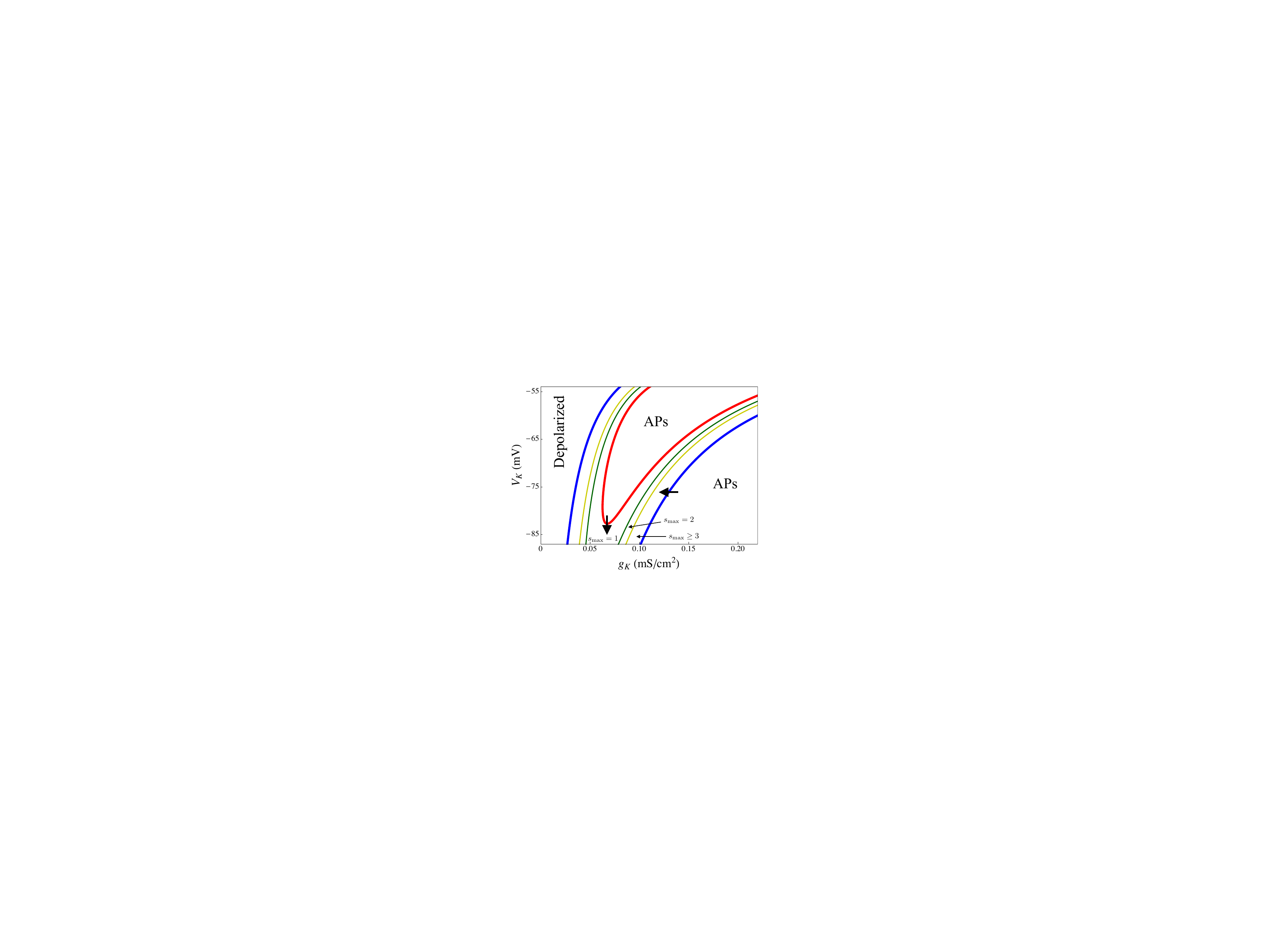}
\caption{Genericity of canard-induced EADs. The $(g_K,V_K)$ parameter plane has been partitioned according to the properties of the folded singularity. Folded nodes and EADs exist in the region enclosed by the blue ($\mu=0$) curves and the red ($\mu=1$) curve. Within this region, the maximal number of EADs that can be observed increases as the parameters are moved from the red $\mu = 1$ boundary to the blue $\mu=0$ boundaries. The two thick arrows indicate possible effects of drugs  that reduce the \Kp\ current conductance (leftward arrow) or increase the magnitude of the \Kp\ Nernst potential (downward arrow). }
\label{fig:twoparam}
\end{figure}

There are two special cases, $\mu = 0$ and $\mu=1$, where the folded node ceases to exist and hence the canard-induced EADs are eliminated. 
The resonance $\mu = 1$ corresponds to the boundary where the folded node becomes a folded focus.
Folded foci do not possess any canards.
Hence, the $\mu=1$ resonance serves as the transition between regular $1^0$ APs and APs with EADs. This is illustrated in a two-parameter diagram, where the conductance of the \Kp\ current ($g_K$) and the \Kp\ Nernst potential ($V_K$) are varied and the asymptotic state of the system \eqref{eq:model} is shown (Fig. \ref{fig:twoparam}). The red curve is the set of parameter values for which $\mu=1$. For parameter values within the region enclosed by the red curve the folded singularity is a folded focus, so only APs are produced (without EADs).   

The dark green curves in Fig. \ref{fig:twoparam} are parameter combinations such that $\mu=1/3$, so in the region delimited by these curves and the red $\mu=1$ curve there is a single maximal canard ($s_{\rm max}=1$) and APs with a single EAD are possible. On the olive curves  $\mu=1/5$ and in the region delimited by these curves and the dark green curves APs with two EADs are possible. This process can be continued to higher odd integer values of $\mu^{-1}$; in the region between the olive curves and blue curves APs with three or more EADs are possible.

On the blue curves $\mu = 0$. The $\mu = 0$ resonance is known as a folded saddle-node (FSN) bifurcation and can occur in several distinct ways.  
The FSN bifurcation of type II (FSN II) is a bifurcation of the desingularized system in which a folded singularity and an ordinary singularity coalesce and swap stability in a hybrid transcritical bifurcation \cite{Guckenheimer2008,Krupa2010}. 
That is, for $\mu>0$, the folded singularity on $L^+$ is a folded node and the ordinary singularity on $S_0^r$ is a saddle equilibrium. For $\mu<0$, the folded singularity on $L^+$ is a folded saddle and the ordinary singularity has moved to $S_0^{a,+}$ where it is a stable node.
Hence, the FSN II bifurcation corresponds to the transition between EADs and stable depolarized steady states (Fig. \ref{fig:twoparam}; left blue curve). 

The other way in which the FSN bifurcation can occur in the desingularized system is via a true transcritical bifurcation of folded singularities. 
That is, for $\mu>0$, there is a folded node on $L^+$ and there is a folded saddle on the $V$-axis. 
At $\mu = 0$, the folded node and folded saddle coalesce, and for $\mu <0$, the folded singularity on $L^+$ is a folded saddle whereas the folded singularity on the $V$-axis is a folded node. 
The slow flow around the folded node on the $V$-axis is directed away from the $V$-axis, and so EADs will not be observed. 
Thus, for $\mu<0$, orbits of the slow flow encounter regular fold points on $L^+$, and the corresponding rhythm exhibits regular APs (without EADs). 
Hence, this FSN bifurcation corresponds to the transition between EADs and regular APs (Fig. \ref{fig:twoparam}; right blue curve). 
We note that, to the best of our knowledge, this type of FSN bifurcation (in which a pair of folded singularities undergo a true transcritical bifurcation) has not yet been reported or studied. 

The two-parameter diagram (Fig. \ref{fig:twoparam}) illustrates that, in this model, there is a large set of $g_K$, $V_K$ parameters in which EADs can be produced. Thus, the behavior is generic, not limited to small regions of parameter space. It also illustrates the precision that GSPT provides in the determination of when EADs are possible, and the maximum number of EADs that are possible.  Finally, the diagram shows that decreasing the \Kp\ conductance, as is done with drugs like azimilide that act as \Kp\ channel antagonists, can induce EADs (thick leftward arrow). Also, increasing the magnitude of the \Kp\ Nernst potential, as in hypokalemia, can induce EADs (thick downward arrow). These observations are consistent with experimental studies \cite{Madhvani2011,Sato2010,Yan2001}.

\section*{Periodic Stimulation \& Mixed-Mode Oscillations}		\label{sec:global}

We have established that EADs originate from canard dynamics around a folded node, and that the canards organize the EADs in both phase and parameter space. 
In this section, we restore the periodic stimulation and study the stimulus-driven EAD attractors. 
Our aim is to explain the bifurcation diagram in Fig. \ref{fig:restitution}.
We will show that the variety of AP morphologies exhibited under various PCLs can be explained by the canards. 

\subsection*{High- and low-frequency pacing: canard-induced mixed-mode oscillations} 	\label{subsec:mmos}

\begin{figure}[ht!]
\centering
\includegraphics[width=5in]{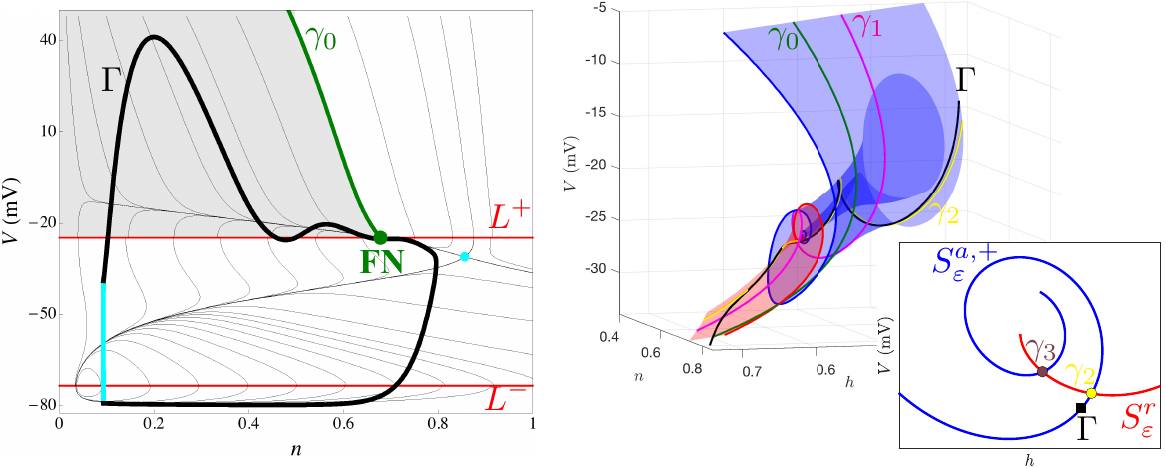}
\put(-362,135){(a)}
\put(-192,135){(b)}
\caption{Geometric mechanism for the stimulus-driven $1^2$ attractor $\Gamma$ (thick black and cyan). Parameters are as in Fig.~\ref{fig:restitution}(a). (a) Comparison of $\Gamma$ with the slow subsystem flow (thin black) and fast subsystem geometric structures. The stimulus (cyan segment) induces a transition from the hyperpolarized sheet to the funnel of the folded node on the depolarized sheet. (b) Comparison of $\Gamma$ with the slow manifolds; $\Gamma$ lies in the sector bounded by the canards $\gamma_1$ and $\gamma_2$, and thus has 2 EADs.}
\label{fig:highfrequency}
\end{figure}

Recall that when there is periodic stimulation, $I_{\rm sti}$, the system entrains to the driving oscillator. 
For low PCLs (i.e., high-frequency pacing), the attractor is a $1^2$ AP with EADs (see Figs. \ref{fig:restitution}(a) and (e)).
Using the results of our geometric analysis from the `\nameref{sec:local}' section above, we now deconstruct the $1^2$ rhythm (Fig. \ref{fig:highfrequency}) and find that it consists of 
\begin{enumerate}[(i)]
\setlength{\itemsep}{0pt}
\item canard-induced EADs around the folded node due to twisted slow manifolds, 
\item a fast transition from the depolarized folded node region to the hyperpolarized slow manifold, and 
\item a stimulus-driven transition from the hyperpolarized slow manifold to the depolarized slow manifold. 
\end{enumerate}
A representative example is shown in Fig. \ref{fig:highfrequency}, where we compare the $1^2$ attractor to the slow and fast subsystems (panel (a)) and to the twisted slow manifolds (panel (b)). 
Note in Fig.~\ref{fig:highfrequency}(a) that the weak canard is approximately given by the stable manifold of the (cyan) saddle, and that the EADs are centered on this weak canard (i.e., the weak canard is the axis of rotation).

The periodic stimulus provides the mechanism for returning orbits to the neighbourhood of the folded node. 
More specifically, the stimulus switches `on' during the slow hyperpolarized segment of the trajectory. 
This drives the orbit away from the hyperpolarized sheet before it can reach the lower firing threshold $L^-$. 
Moreover, the amplitude of the stimulus pulse is large enough that it pushes the orbit past the repelling sheet of the critical manifold and into the basin of attraction of the depolarized sheet, $S_{\eps}^{a,+}$.
The timing of the stimulus is also such that the orbit is injected into the rotational sector enclosed by the maximal canards $\gamma_1$ and $\gamma_2$, and hence exhibits 2 EADs. 
This combination of a local canard mechanism (for the EADs) and a global (stimulus-induced) return mechanism is known as a canard-induced mixed-mode oscillation (MMO) \cite{Brons2006}.  

\begin{figure}[ht!]
\centering
\includegraphics[width=5in]{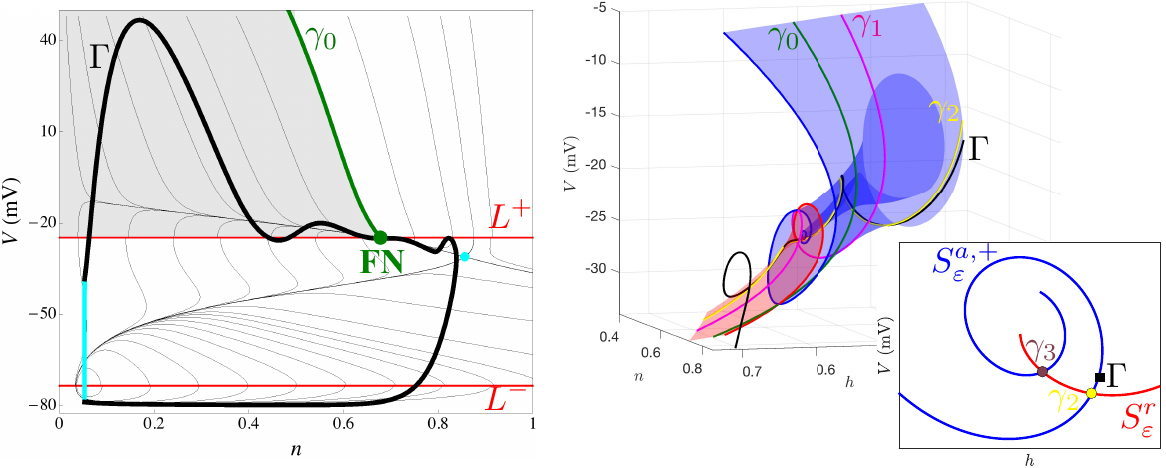}
\put(-362,135){(a)}
\put(-192,135){(b)}
\caption{Local and global mechanisms for a stimulus-driven $1^3$ attractor $\Gamma$ (thick black and cyan). Parameters are as in Fig.~\ref{fig:restitution}(d). (a) Comparison of $\Gamma$ with the slow subsystem flow (thin black) and fast subsystem geometric structures in the $(V,n)$ projection. The stimulus (cyan segment) projects the orbit into the funnel of the folded node on the depolarized sheet. (b) The orbit is injected into the rotational sector delimited by the canards $\gamma_2$ and $\gamma_3$, and hence exhibits 3 EADs.}
\label{fig:lowfrequency}
\end{figure}

Similarly, we find that for large PCLs (i.e., low-frequency pacing), the stimulus-driven $1^3$ attractor is a canard-induced MMO with period set by the PCL (see Fig. \ref{fig:restitution}(d) and (e)). The $1^3$ MMO attractor consists of (local) canard-induced EADs around the folded node combined with a global stimulus-driven return that projects orbits from the hyperpolarized sheet into the rotational sector enclosed by the canards $\gamma_2$ and $\gamma_3$ (Fig. \ref{fig:lowfrequency}). 

\subsection*{Intermediate-frequency pacing: EAD alternans due to reinjection into adjacent rotational sectors} 	\label{subsec:alternans}
\begin{figure}[ht!]
\centering
\includegraphics[width=5in]{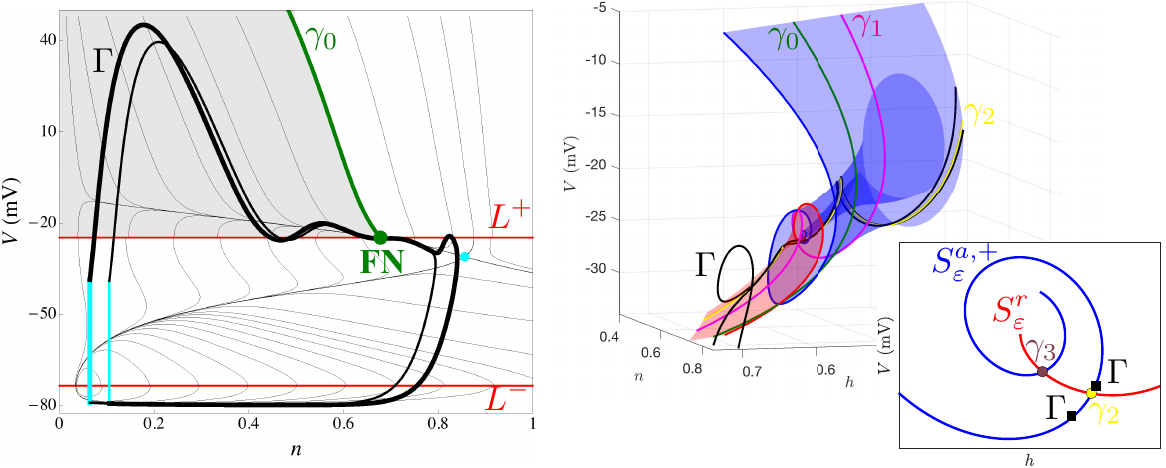}
\put(-362,135){(a)}
\put(-192,135){(b)}
\caption{Geometric mechanism for stimulus-driven $1^2 1^3$ alternans $\Gamma$. Parameters are as in Fig.~\ref{fig:restitution}(c). (a) Comparison of $\Gamma$ with the slow flow (thin black). The stimulus (cyan) projects the orbit into the funnel at different locations, causing $\Gamma$ to visit different rotational sectors. (b) The orbit alternately enters the rotational sector enclosed by $\gamma_1$ and $\gamma_2$ (2 EADs), and the rotational sector enclosed by $\gamma_2$ and $\gamma_3$ (3 EADs).}
\label{fig:intfreqalternans}
\end{figure}

In Fig.~\ref{fig:restitution}(e), we found that there is a band of intermediate pacing frequencies for which the stimulus-driven attractor is a $1^2 1^3$ alternator (see Fig.~\ref{fig:restitution}(b) for a representative time series).  
We now compare the $1^2 1^3$ attractor with the underlying geometric structures of the model cell (Fig.~\ref{fig:intfreqalternans}). 
As in the low- and high-frequency forcing cases, we find that the $1^2$ and $1^3$ segments are each canard-induced MMOs. 
The difference here is that the timing of the stimulus is such that the orbit visits different (contiguous) rotational sectors on each stimulus pulse. 

The $1^2 1^3$ alternans attractor, $\Gamma$, decomposes as follows.
Starting on the hyperpolarized sheet, the first stimulus pulse (Fig. \ref{fig:intfreqalternans}(a); leftmost cyan segment) projects the orbit $\Gamma$ into the rotational sector enclosed by $\gamma_2$ and $\gamma_3$ (Fig. \ref{fig:intfreqalternans}(b); inset -- black marker above $\gamma_2$). 
Thus, $\Gamma$ exhibits 3 EADs. 
After these 3 EADs are completed, the orbit transitions to the hyperpolarized sheet where it slowly drifts towards the firing threshold $L^-$. Before it can reach $L^-$, the next stimulus pulse (Fig. \ref{fig:intfreqalternans}(a); rightmost cyan segment) projects the orbit into the rotational sector enclosed by $\gamma_1$ and $\gamma_2$ (Fig. \ref{fig:intfreqalternans}(b); inset -- black marker below $\gamma_2$), and thus $\Gamma$ exhibits only 2 EADs. 
The orbit then returns to the hyperpolarized sheet where it again slowly drifts towards $L^-$.
Since $\Gamma$ only underwent 2 EADs, the APD is shorter (compared to the previous one) and the corresponding diastolic interval (DI) is longer.
As such, the orbit is able to drift further along the hyperpolarized sheet before the next stimulus occurs.
Once the stimulus `switches on', the process repeats periodically, thus producing the $1^2 1^3$ attractor.  

\subsection*{Intermediate-frequency pacing: dynamical chaos and intermittency due to sensitivity near maximal canards} 	\label{subsec:chaos}

\begin{figure}[ht!]
\centering
\includegraphics[width=5in]{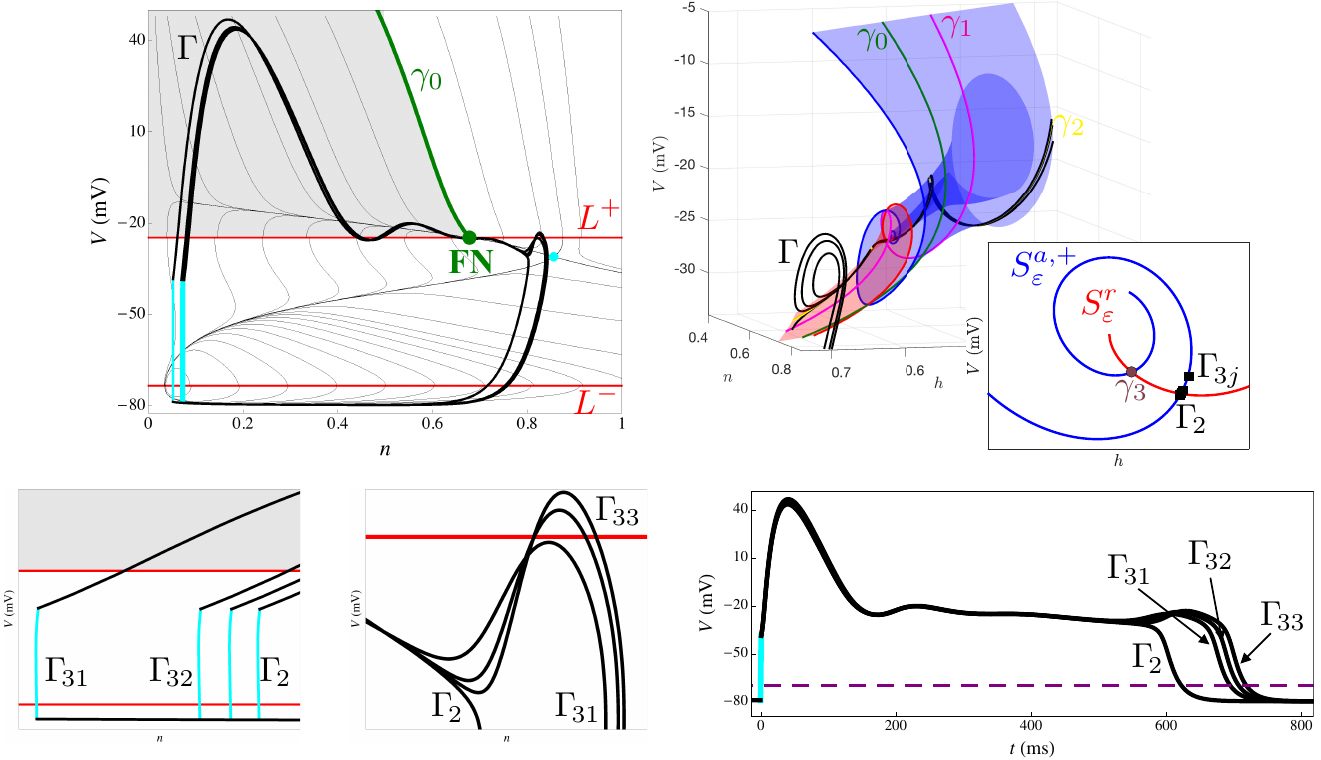}
\put(-340,198){(a)}
\put(-188,198){(b)}
\put(-370,66){(c)}
\put(-276,66){(d)}
\put(-176,66){(e)}
\caption{Geometric explanation of `dynamical chaos' and intermittency. 
The cause for the variability in the number and magnitude of the EADs is that $\Gamma$ peels off $\gamma_2$ at different times.
Parameters are as in Fig.~\ref{fig:restitution}(d). 
(a) Projection of the $1^2 (1^3)^3$ attractor, $\Gamma$, into the $(V,n)$ plane. 
(b) The orbit stays close to the maximal canard $\gamma_2$ on each return to $S_{\eps}^{a,+}$.
(c) Zoom of the $(V,n)$ plane where the stimuli are applied. 
(d) Zoom of the EADs as they peel off the maximal canard $\gamma_2$.
(e) Time series of $\Gamma_2$ and $\Gamma_{3j}$ for $j=1,2,3$.
The APD (DI) is the time spent above (below) the threshold $V = -70~$mV (dashed purple).
}
\label{fig:intfreqchaos}
\end{figure}
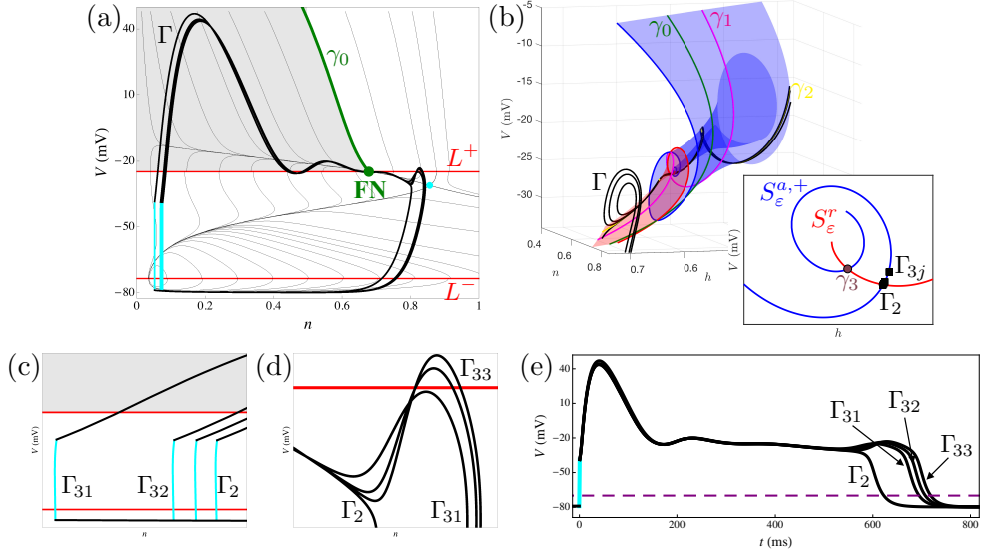

In Fig.~\ref{fig:restitution}(e), we found a band of intermediate pacing frequencies for which the model cell exhibited seemingly chaotic and intermittent behaviour. 
Here, we show that the complex EAD signatures arise from the crossing of a maximal canard. 
We do this for a representative $1^2 (1^3)^3$ attractor (Fig.~\ref{fig:intfreqchaos}), which we denote by $\Gamma$. 
As before, the individual APs with EADs are canard-induced MMOs. 
The variability in the number and magnitude of the EADs is due to the stimulus, which perturbs the orbit away from the hyperpolarized sheet at different locations on each pulse.

Let $\Gamma_2$ denote the $1^2$ segment of $\Gamma$, and $\Gamma_{3,j}, j=1,2,3$ denote the $1^3$ segments of $\Gamma$, i.e., $\Gamma = \Gamma_2 \cup \Gamma_{31} \cup \Gamma_{32} \cup \Gamma_{33}$. 
Starting with $\Gamma_2$, the stimulus (Fig.~\ref{fig:intfreqchaos}(c); cyan) induces a fast transition to the depolarized sheet close to $\gamma_2$ and in the sector between $\gamma_1$ and $\gamma_2$ (Fig.~\ref{fig:intfreqchaos}(b)), and hence there are 2 EADs.
The intrinsic dynamics of the model cell return $\Gamma_2$ to $S_{\eps}^{a,-}$ where it slowly drifts to smaller $n$. 
The next stimulus initiates $\Gamma_{31}$ and causes the orbit to enter the depolarized phase in the rotational sector bound by $\gamma_2$ and $\gamma_3$. 
The additional EAD produced in $\Gamma_{31}$ extends the APD compared to that of $\Gamma_2$ (Fig.~\ref{fig:intfreqchaos}(e)).
As such, the DI of $\Gamma_{31}$ is shorter than that of $\Gamma_2$.  
This means $\Gamma_{32}$ is initiated on $S_{\eps}^{a,-}$ at a larger $n$ value (Fig.~\ref{fig:intfreqchaos}(c)), and enters $S_{\eps}^{a,+}$ closer to $\gamma_2$.
Since $\Gamma_{32}$ follows $\gamma_2$ more closely than $\Gamma_{31}$, (i) the resulting EADs are larger amplitude (Fig.~\ref{fig:intfreqchaos}(d)), (ii) the APD is longer, and (iii) the DI is shorter.
Consequently, $\Gamma_{33}$ is initiated on $S_{\eps}^{a,-}$ at a larger $n$ value, enters $S_{\eps}^{a,+}$ closer to $\gamma_2$, and hence exhibits the (i) largest EADs, (ii) longest APD, and (iii) shortest DI. 

The other complex MMO signatures reported in Fig.~\ref{fig:restitution}(e) emerge by the same mechanism.
That is, the $(1^2)^p (1^3)^q$ attractors for $p,q \in \mathbb{N}$ arise because the PCL is such that the orbit enters the depolarized sheet close to the maximal canard $\gamma_2$.
Since the behaviour of trajectories near a maximal canard is exponentially sensitive \cite{Wexy2005}, small changes in the PCL manifest as significant changes in the number, amplitude, and duration of EADs on each pulse. Likewise, at such PCL values, small changes in initial conditions have large effects on the $V$ time course, the hallmark of chaos.

\section*{Discussion}		\label{sec:discussion}

It has been demonstrated previously that early afterdepolarizations produced by a simple cardiomyocyte model \cite{Sato2010}, a reduction of the Luo-Rudy 1 model \cite{Luo1991}, are the consequence of carnard dynamics in the vicinity of a folded node singularity \cite{kugler2018}, a result further illuminated through the geometric analysis shown in Figs. \ref{fig:slowflow}--\ref{fig:sectors}. We showed that these dynamics are robust in the $(g_K,V_K)$ parameter plane (Fig. \ref{fig:twoparam}). These parameters were chosen since they can be modulated by drugs or environment; $g_K$ is reduced by \Kp\ channel antagonists such as azimilide, while $V_K$ is increased in magnitude in hypokalemia. Figure \ref{fig:twoparam} predicts that both manipulations can induce EADs, and indeed both manipulations have been shown to do this in experiments \cite{Madhvani2011,Sato2010,Yan2001}. 

The second set of results from our study involves the paced system, which receives periodic depolarizing stimuli (Fig. \ref{fig:restitution}). Each stimulus pushes an orbit into the basin of attraction of the depolarized attracting sheet, triggering an action potential that can be a mixed-mode oscillation if EADs are produced. For high- and low-frequency pacing, orbits land in the rotational sectors delimited by the maximal canards and stay far from any of the maximal canards, so that the voltage time course exhibits regular, periodic behavior. At high stimulus frequencies, the orbits land in the rotational sector with 2 EADs, so each AP is a mixed-mode oscillation with 2 EADs. At low stimulus frequencies, the orbits land in the rotational sector with 3 EADs, so each AP is a mixed-mode oscillation with 3 EADs. The number of EADs depends upon the rotational sector in which the orbit lands in response to the stimulus (Figs. \ref{fig:highfrequency}--\ref{fig:intfreqchaos}).  The EAD alternans observed for intermediate-frequency pacing emerge because the stimulus current alternately projects the orbit into different rotational sectors on each pulse. In some cases, the outcome can be quite complex, with a sequence of mixed-mode oscillations of different durations and numbers of EADs. This behavior is what was referred to as ``dynamical chaos" in earlier publications \cite{Sato2010,Tran2009}.

The advantage of the minimal model for the analysis presented here is its low dimensionality. More realistic cardiomyocyte models can have 40 or more dimensions, reflecting many types of ionic currents and in many cases equations for \Ca\ handling in the cytosol, the sarcoplasmic reticulum (SR), and the subspace between the SR and the cell membrane \cite{Luo1994b,Luo1994a,Kurata2005,Ohara2011,Williams_sb2010}.  One major advantage of these larger models is that they have more biological detail that allows for simulation of, for example, the application of pharmacological agents that act as antagonists for specific types of ion channels, such as inward-rectifying \Kp\ channels, while the minimal model incorporates only a single type of \Kp\ current and a single type of \Ca\ current. With the correct parameterizations, these more complete models are capable of reproducing the various forms of EADs that have been characterized, each with different, but partially overlapping, biophysical mechanisms  \cite{Antzelevitch2011}, while the minimal model was developed to produce EADs of a particular type. EADs are divided broadly into types according to the timing of the events: ``phase-2 EADs" occur during the plateau of an elongated AP, and ``phase-3 EADs" occur during the falling phase of the AP.  There are also ``depolarizing afterdepolarizations" that occur after the completion of the action potential. The analysis that we performed herein on a minimal model suggests that the dynamics underlying some phase-2 EADs are canard induced, and we speculate that this will be the case in more complete biophysical models. While the full geometric singular perturbation analysis done with the minimal model is not possible with the high-dimensional models, it is possible to perform a less complete analysis, such as determining the existence of folded node singularities. Indeed, such an analysis is important for establishing that canard dynamics are the basis of phase-2 EADs in more complete models, and is currently being undertaken by our group. 

Why does it matter whether EADs are due to canard dynamics near a folded node singularity? Although it sounds very abstract, the ramifications of knowing this can be very important and useful. As we have demonstrated, if the EADs are associated with a folded node singularity, then one can simply analyze the eigenvalues of the reduced desingularized system at the folded node to determine how many EADs are possible. Also, through analysis of the eigenvalues, one can determine parameter changes that will enhance EAD production or eliminate the EADs. In particular, one can determine regions of parameter space where canard-induced EADs are not possible, without the need to perform any numerical integrations (as in Fig. \ref{fig:twoparam} and \cite{kugler2018}). So once EADs are linked to folded node singularities, one gains a great deal of predictive capability. In addition to this, knowing the dynamical mechanism for the EADs helps in the understanding of complex behavior, such as dynamical chaos, that would be hard or impossible to understand from the viewpoint of interacting ionic currents (i.e., a biophysical interpretation).  Knowing which ion channels are key players in EADs is of course important, and can provide targets for pharmacological or genetic manipulation, but the complexity of the multiscale nonlinear dynamical system provides limitations to interpreting behavior without mathematical tools such as GSPT. 

The theory of folded singularities has been applied to numerous biological systems. This includes intracellular \Ca\ dynamics \cite{Harvey2011}, the electrical activity of neurons \cite{Rotstein2008,Rubin2007,Rubin2008} and pituitary cells \cite{Vo2013}, and mixed-mode oscillations that are likely canard-induced have been observed in the oxidation of platinum \cite{Krischer1992}, dusty plasmas \cite{Mikikian2008}, and chemical oscillations \cite{Petrov1992,Rotstein2003}. The demonstration that some forms of EADs are canard-induced, at least in a minimal cardiomyocyte model, adds cardiac cells to the growing list of the biological and chemical systems whose dynamics are organized by folded singularities. Our system is novel, however, in that it is periodically forced under normal (i.e., physiological) conditions, where the forcing is initiated at the sinoatrial node. As we demonstrated here, this forcing can lead to complicated dynamics due to the injection of the orbit into different rotational sectors, so that the number of EADs produced following each stimulus can vary. The result can appear to be unpredictable, and chaotic, and sensitive to small changes in the forcing frequency and initial conditions. Whether this complex behavior is exhibited in a physiological setting, within an intact heart, is unclear. It is generally accepted that EADs can lead to arrythmias \cite{Cranefield1991,Lerma2007stochastic,Shimizu1997,Shimizu1991}, including ventricular tachycardia, but it has not been establshed that complex, chaotic behavior at the single myocyte level contributes to this. 

\section*{Conclusions}		\label{sec:conclusions}

In this report, we showed the benefits of a 2-slow/1-fast analysis of a model for cardiac early afterdepolarizations. Knowing that the small EADs are due to canards organized around a folded node singularity not only explains the origin of the EADs, but provides a viewpoint through which one can comprehend important behaviors. For example, an analysis of the eigenvalues of the folded-node singularity provides information on the number of EADs that are possible for different parameter sets. It also explains why inhibition of \Kp\ channels or a hypokalemic environment facilitates EAD production. Finally, it provides a solid basis for understanding the effects of periodic stimulation of cardiomyocytes. We used this technique to show why more EADs are generated at low-frequency pacing than at a higher pacing frequency. The technique was also used to explain the origin of complex alternan behavior that occurs with intermediate-frequency pacing. Overall, the use of slow-fast analysis provides information on the dynamics of this multi-timescale system that are hard or impossible to comprehend from a purely biophysical analysis (i.e., in terms of the effects of different ionic currents) or from computer simulations alone.

\small
\bibliography{version5}

\end{document}